\PassOptionsToPackage{maxbibnames=99}{biblatex}
\documentclass[a4paper, oneside, twocolumn, notitlepage, 10pt]{extarticle_ecoc}
\usepackage{ecoc}
\usepackage{orcidlink}

\usepackage{amsmath, amssymb, dsfont, mathtools}
\usepackage[nolist]{acronym}
\usepackage[labelformat=simple]{subcaption}

\usepackage{placeins}

\usepackage{gensymb}
\usepackage{tikz}
\acrodef{IAB}{Integrated Access and Backhaul}
\acrodef{SBS}{Small Base Station}
\acrodef{MBS}{Macro Base Station}
\acrodef{BS}{Base Station}
\acrodef{UE}{User Equipment}
\acrodef{LoS}{Line-of-Sight}
\acrodef{NLoS}{Non-Line-of-Sight}
\acrodef{FSO}{free space optics}
\acrodef{SINR}{signal-to-interference-plus-noise ratio}
\acrodef{CU}{cost units}
\acrodef{EU}{energy units}

\begin{document}
\selectlanguage{english}    

\setlength{\belowdisplayskip}{2pt} 
\setlength{\belowdisplayshortskip}{2pt}
\setlength{\abovedisplayskip}{2pt} 
\setlength{\abovedisplayshortskip}{2pt}
\setlength{\textfloatsep}{12pt plus 2pt minus 2pt}
\setlength{\dbltextfloatsep}{10pt plus 2pt minus 2pt}


\title{Energy-Efficient FSO Reconfiguration under User Mobility in Hybrid Fiber-IAB Backhaul}


\author{Piotr Lechowicz~\orcidlink{0000-0003-2555-5187}~\textsuperscript{*,1}, Charitha Madapatha ~\orcidlink{0000-0003-3364-282X}~\textsuperscript{1}, Carlos Natalino~\orcidlink{0000-0001-7501-5547}~\textsuperscript{1}, Tommy Svensson~\orcidlink{0000-0002-0521-3107}~\textsuperscript{1}, \\Paolo Monti~\orcidlink{0000-0002-5636-9910}~\textsuperscript{1}}

\maketitle                  


\begin{strip}
    \begin{author_descr}
        
        Department of Electrical Engineering, Chalmers University of Technology, Gothenburg, Sweden \\
        \textsuperscript{(*)}
        \textcolor{blue}{\uline{piotr.lechowicz@chalmers.se}}
        
    \end{author_descr}
    \vspace{0.5cm}
\end{strip}


\begin{strip}
    \begin{ecoc_abstract}
        User mobility creates stochastic, time-varying backhaul demand that static capacity provisioning cannot match. 
        We propose a closed-loop, load-aware hysteresis controller for hybrid fiber-IAB-FSO backhaul and show that energy drops faster than coverage: 8\% to 44\% energy savings cost only 0.9\% to 6.7\% coverage.
        \copyright{}~2026 The Author(s)  \\
        This is the authors' version of this publication.
    \end{ecoc_abstract}
    
\end{strip}

\section{Introduction}
Future 6G networks will require massive densification of low-power \acp{SBS} for high data rates and coverage.
Since providing fiber backhauling to every \ac{BS} is prohibitively expensive, hybrid architectures combining fiber, \ac{IAB}, and \ac{FSO} have emerged as a practical alternative \cite{2024_Raj_EfficientConfluentEdge, 2016_Zhang_FiberWirelessIntegrated, 2025_Madapatha_JointFiberFree}.
In standard in-band \ac{IAB} deployments, high-power \acp{MBS} act as fiber-connected donor nodes, sharing access spectrum to provide wireless backhaul to intermediate \ac{IAB}-child nodes (\acp{SBS}).
Deploying \ac{FSO} as a high-capacity complement to \ac{IAB} and a cost-effective substitute for fiber yields substantial savings \cite{2025_Madapatha_JointFiberFree} while mitigating the bandwidth starvation inherent in pure in-band \ac{IAB} deployments \cite{2020_Polese_IntegratedAccessBackhaul}.
In such multi-technology deployments, a confluent controller \cite{2024_Raj_EfficientConfluentEdge} can use instantaneous wireless traffic data in a closed loop to dynamically configure both radio and optical links. 

Conventional network planning assumes static user distributions, disregarding that \ac{UE} mobility causes spatial load fluctuations and produces time-varying backhaul demand. 
The set of \acp{SBS} that actually need high-capacity backhaul shifts stochastically with traffic, so no static \ac{FSO} assignment can match the demand profile at all times. 
Keeping links active during low-demand intervals wastes energy; keeping them inactive during surges costs coverage. 
Recent physical-layer optical-to-mmWave switching demonstrations target weather resilience \cite{2025_Vargemidou_FieldTrialSDNControlled, 2025_Kyriazi_DemonstrationSDNenabledResilient}, yet their use for system-wide coverage maximization and load-driven energy reduction remains unaddressed.
While dynamic \ac{FSO} reconfiguration under user mobility is explored in flying-platform-based networks \cite{2018_Gu_NetworkTopologyReconfiguration}, it has not been investigated in confluent scenarios operating alongside \ac{IAB}.
Although \ac{IAB} topology design is studied under static \cite{2025_Madapatha_JointFiberFree, 2025_Zhang_Optimizing6GDensea, 2023_Huang_BayesianApproachDesign, 2021_Madapatha_TopologyOptimizationRoutinga} or mobile user scenarios \cite{2026_Lechowicz_UserMobilityAwareOptimizationFiber}, the mobility-aware dynamic operation of hybrid fiber-\ac{IAB}-\ac{FSO} topologies remains an open challenge.

We address this gap with a closed-loop heuristic that activates auxiliary \ac{FSO} links only when short-term average per-\ac{SBS} user load exceeds a threshold, with hysteresis suppressing switching transients near the decision boundary. To our knowledge, this is the first load-driven reconfiguration scheme for hybrid fiber-\ac{IAB}-\ac{FSO} backhaul evaluated under realistic \ac{UE} mobility conditions.
Our results reveal a favorable asymmetry: energy drops faster than coverage. 
Capping coverage loss at 0.9\% yields 8\% energy savings, while accepting a 6.7\% drop yields up to 44\%.

\section{Network architecture}

\begin{figure*}[t]
    \centering    
    \begin{subfigure}[t]{.29\textwidth}
        \includegraphics[width=\textwidth]{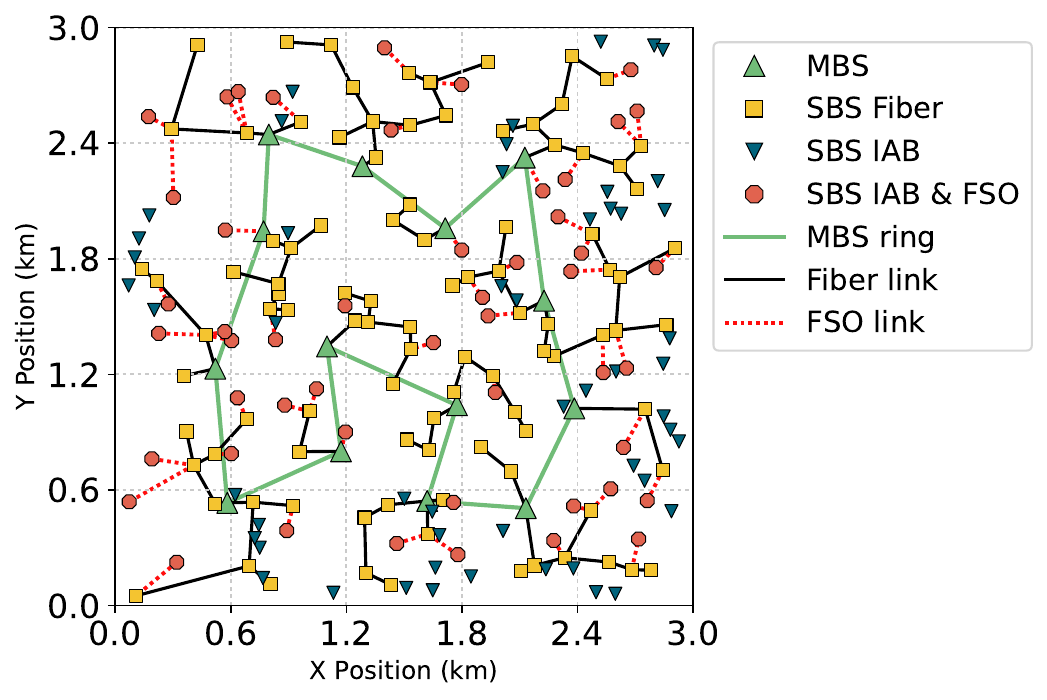}
        \caption{Network topology.}
        \label{fig:example:topology}
    \end{subfigure}\hfill
    \begin{subfigure}[t]{.28\textwidth}
        \includegraphics[width=\textwidth]{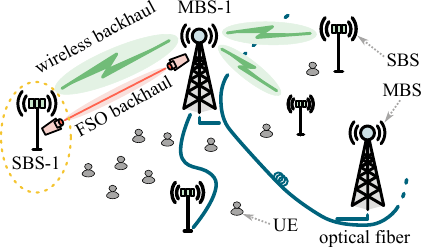}
        \caption{Confluent backhaul between SBS-1 and MBS-1.}
        \label{fig:example:fso-backhaul}
    \end{subfigure}\hfill
    \begin{subfigure}[t]{.36\textwidth}
        \includegraphics[width=\textwidth]{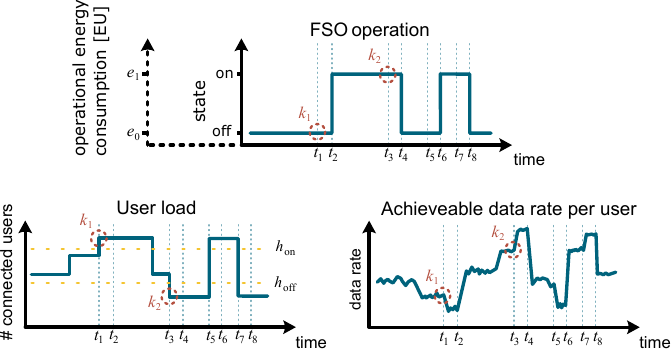}
        \caption{Operational values of SBS-1.}
        \label{fig:example:plots}
    \end{subfigure}%
    \vspace{.3cm}
    \caption{Illustrative example of hybrid topology operation.}
    \label{fig:example}
    \vspace{.3cm}
\end{figure*}

We consider a hybrid deployment in which fiber-connected \acp{MBS} (\ac{IAB}-donors) provide wireless backhaul to \acp{SBS} (i.e., \ac{IAB}-child nodes) that collectively serve mobile \acp{UE} over a metropolitan area.
Fig.~\ref{fig:example:topology} illustrates a topology where 50\% of \acp{SBS} are fiber-backhauled (yellow squares), 25\% rely on \ac{IAB} (blue triangles), and 25\% employ confluent \ac{IAB} with auxiliary \ac{FSO} backhaul (red circles). 
The confluent backhaul between two nodes is depicted in Fig.~\ref{fig:example:fso-backhaul}, where \ac{IAB}-child node SBS-1 communicates with \ac{IAB}-donor MBS-1 via either \ac{IAB} or \ac{FSO}.
Nodes are equipped with dynamic switching capabilities between the optical and mmWave domains \cite{2025_Vargemidou_FieldTrialSDNControlled, 2025_Kyriazi_DemonstrationSDNenabledResilient}.
Fig.~\ref{fig:example:plots} presents the operational conditions of SBS-1. 
Operating as a closed-loop system, a centralized controller exploits the monitored wireless load to actively toggle the optical \ac{FSO} state: exceeding $h_{\mathrm{on}}$ activates the \ac{FSO} link, while dropping below $h_{\mathrm{off}}$ triggers deactivation.
The \ac{FSO} link operates in two energy states: $e_0$ \ac{EU} (standby) and $e_1$ \ac{EU} (active).
In this example, exceeding $h_{\mathrm{on}}$ at $t_1$ (event $k_1$) activates the \ac{FSO} link with a slight delay at $t_2$, necessary to configure the link, align the \ac{FSO} antennas, and stabilize the channel.
The per-user data rate initially drops due to increasing load, but recovers once the \ac{FSO} backhaul is activated.
Conversely, the load drops below $h_{\mathrm{off}}$ at $t_3$ (event $k_2$), deactivating \ac{FSO} at $t_4$.
The per-user data rate temporarily rises due to reduced load before decreasing upon \ac{IAB} fallback.

To formalize the topology, let $\mathbf{D}$, $\mathbf{C}$, $\mathbf{S}$, and $\mathbf{U}$ denote the sets of \ac{IAB}-donors, child nodes, fiber-backhauled \acp{SBS}, and \acp{UE}, respectively.
Let $\mathbf{F} \subset \mathbf{C}$ be the subset of child nodes equipped with \ac{FSO}.
We define $z_f \in \{0,1\}$ as the binary operational state of node $f \in \mathbf{F}$ (1 = \ac{FSO} active, 0 = \ac{IAB} fallback). 
For \ac{UE} $u$, $b_u$ denotes the serving \ac{BS} and $\mathrm{SINR}_u$ the experienced \ac{SINR}.
Let $D(c)$ denote the donor serving child node $c$.
For any $c \in \mathbf{C}$, $\mathrm{SINR}_c$ is the backhaul link \ac{SINR} between child $c$ and its donor $D(c)$.
We define $\beta_d \in [0, 1]$ as the access-backhaul sharing ratio for $d \in \mathbf{D}$.
The access bandwidth for node $d \in \mathbf{D}$ is $W_{\mathrm{ac},d}=W \cdot (1 -\beta_d)$.
The remaining spectrum is shared among the associated child nodes proportionally to their load:
\begin{equation}
W_{\mathrm{bh},c} = \frac{\beta_{D(c)} N_{u, c} W}{N_{c, d}},\quad c \in \mathbf{C}
\end{equation}%
where $N_{u,c}$ is the number of \acp{UE} at child $c \in 
\mathbf{C}$, and $N_{c, d}$ is the total \acp{UE} served via all children of donor $d$.
For each child node, the access spectrum is $W_{\mathrm{ac},c} = W - W_{\mathrm{bh},c}$.
We define $\mathrm{SE}_x = \log_2\left(1 + \mathrm{SINR}_x\right)$ as the spectral efficiency. 
The downlink rate $R_u$ for \ac{UE} $u$ is:
{
\setlength{\arraycolsep}{0pt}%
\begin{equation}%
    \resizebox{\columnwidth}{!}{$ %
    \displaystyle %
    R_u = \begin{cases}
        \frac{W_{\mathrm{ac},d}}{N_{u, d}} \mathrm{SE}_u, & b_u = d \in \mathbf{D}\\
        \min \left\{
            \begin{array}{l}
            \frac{W_{\mathrm{ac},x}}{N_{u, x}} \mathrm{SE}_u \\
            \frac{W_{\mathrm{bh},x}}{N_{x, d}} \mathrm{SE}_c
            \end{array}
            \right\},%
            &\begin{array}{ll}%
            b_u \in \:&\{x \in \mathbf{C} \setminus \mathbf{F}\}\:\cup \\
             &\{x \in \mathbf{F}: z_x = 0\}
            \end{array} \\
        \frac{W}{N_{u, x}} \mathrm{SE}_u,& \begin{array}{ll}
             b_u \in \:&\{x \in \mathbf{S}\}\:\cup\\
             &\{x \in \mathbf{F}: z_{x} = 1\}
        \end{array}
    \end{cases}.
    $}%
    \label{eq_data-rate}
\end{equation}
}

The backhaul bandwidth is shared between the \ac{IAB}-donor and its children proportionally to the number of \acp{UE} connected to each child \ac{BS}, and each child equally divides its access bandwidth among associated \acp{UE}.
Specifically, $N_{u,d}$ is the number of \acp{UE} served by donor $d \in \mathbf{D}$, and $N_{u,x}$ is the number of \acp{UE} connected to node $x \in \mathbf{C} \cup \mathbf{S}$.

We formulate a bi-objective problem that maximizes coverage probability $C$ while minimizing energy consumption $E$ under a fixed topology.
Coverage is defined as the average fraction of users over time $T$ achieving a data rate above threshold $\eta$.
The \ac{FSO} operational energy is modeled as a two-state system: standby ($e_0$ \ac{EU}) and active ($e_1$ \ac{EU}), enabling $E$ to be normalized between 0 (all inactive) and 1 (all active).
The specific objectives are defined as:

\begin{equation}
\resizebox{\columnwidth}{!}{$ %
    \displaystyle %
    C = \frac{1}{|\mathbf{U}|\cdot T} \sum_{u \in \mathbf{U}, t \in T} \mathds{I}\left(R_u(t) \geq \eta\right), \qquad E = \frac{1}{|\mathbf{F}|}\sum_{f \in \mathbf{F}} z_f,
    $}%
    \label{eq:objectives}
\end{equation}
where $\mathds{I}(\cdot)$ denotes the indicator function.

We propose a hysteresis-based heuristic \textit{HH} that toggles \ac{FSO} links based on the average load $\bar{N}_{u,f}$ over a monitoring window $\tau$. 
Two thresholds, $h_{\mathrm{on}} > h_{\mathrm{off}}$, create a dead-band that prevents rapid switching transients.
For each node $f \in \mathbf{F}$, the algorithm:
\begin{itemize}
   \item turns \ac{FSO} on if $\bar{N}_{u,f} \geq h_{\mathrm{on}}$, and sets $z_f = 1$,
   \item turns \ac{FSO} off if $\bar{N}_{u,f} \leq h_{\mathrm{off}}$, and sets $z_f = 0$.
\end{itemize}%

Let HH($a,b$) denote the HH algorithm executed with thresholds $h_{\mathrm{on}}=a, h_{\mathrm{off}}=b$.
As baselines, we consider \textit{All On}, where all \ac{FSO} links are permanently active, and \textit{All Off}, where all \ac{FSO} links are permanently deactivated.

\section{Numerical Results}

The topology (Fig.~\ref{fig:example:topology}) comprises 14 \acp{MBS} and 201 \acp{SBS} across a 3~km\textsuperscript{2} area.
Specifically, 100 \acp{SBS} are fiber-backhauled, 50 use \ac{IAB} with auxiliary \ac{FSO}, and the remainder are pure \ac{IAB}-child nodes.
The network operates at 28~GHz with a system bandwidth $W=1$~GHz \cite{2025_Madapatha_JointFiberFree, 2026_Lechowicz_UserMobilityAwareOptimizationFiber}.
Channel characteristics follow the 5GCM Urban Macro (UMa) model \cite{Rappaport_2017_OverviewMilimiterWave} with 3GPP TR~38.901 blocking probability \cite{ETSI2020}.
Transmit powers are 40~dBm for \acp{MBS} and 24~dBm for \acp{SBS}.
All \ac{BS} sites use three-sector antennas (60$\degree$ half-power beamwidth, 24~dBi main-lobe gain, $-$2~dBi side-lobe gain).
We simulate 1000 \acp{UE} using a random waypoint model (1--15~m/s, 0--10~s pauses). 
The 3~km\textsuperscript{2} area is partitioned into a 3$\times$3 grid; \acp{UE} select destinations equally between their current and the central square, inducing temporary congestion in the center of the network.
\Acp{UE} associate with the \ac{BS} offering the highest received power (computed as a product of transmit power, antenna gain, and path loss).

\Ac{FSO} operational energy consumption is normalized between all-active and all-inactive states.
Performance is evaluated over a 1-hour window ($T$) with 1-second granularity.
We set $\tau=30$ seconds to balance responsiveness and stability. 
We empirically select HH thresholds to evaluate the full operational trade-off from conservative (coverage-prioritizing) to aggressive (energy-saving) configurations:
$(h_{\mathrm{on}}, h_{\mathrm{off}}) \in \{(2,1), (3,1), (3,2), (5,2)\}$.
Results are averaged over 10 independent \ac{UE} initial positions and mobility patterns.
To ensure visual clarity, 95\% confidence intervals are omitted from all plots;
the time-averaged (and absolute maximum per-time-step) error margins are 0.3~p.p. (2.7~p.p.) for user coverage and 0.012 (0.06) for normalized \ac{FSO} energy consumption.

\begin{figure}[t]
    \centering
    \includegraphics[width=.95\columnwidth, trim={0 0.2cm 0 0}, clip]{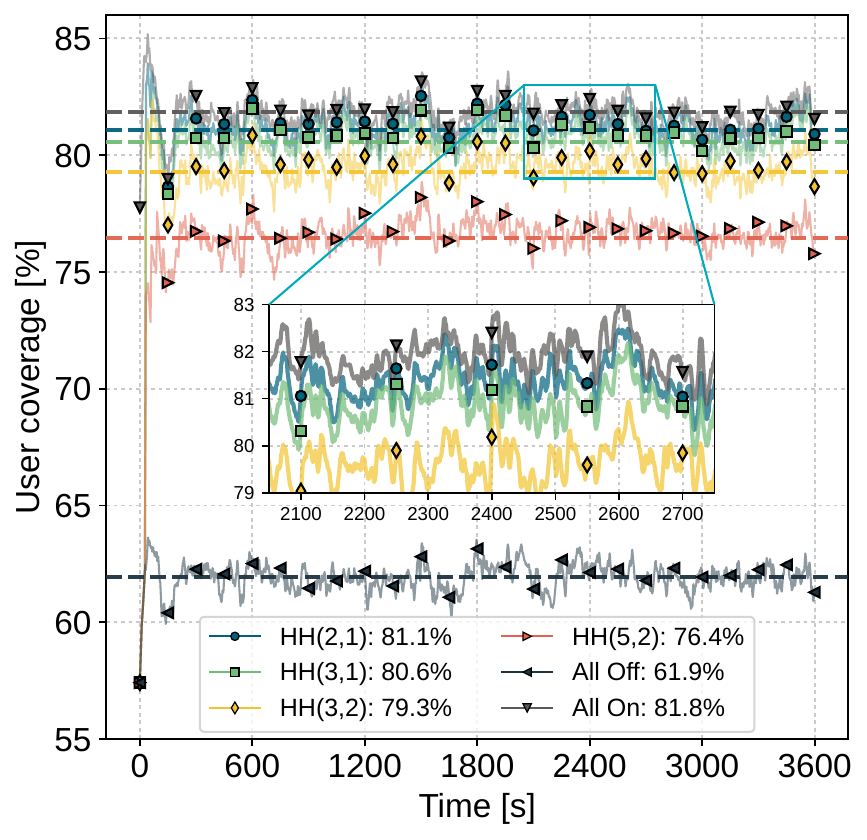}
    \caption{User coverage over time.}
    \label{fig:results-coverage}
\end{figure}

\begin{figure}[t]
\centering
    \includegraphics[width=\columnwidth, trim={0 0.2cm 0 0}, clip]{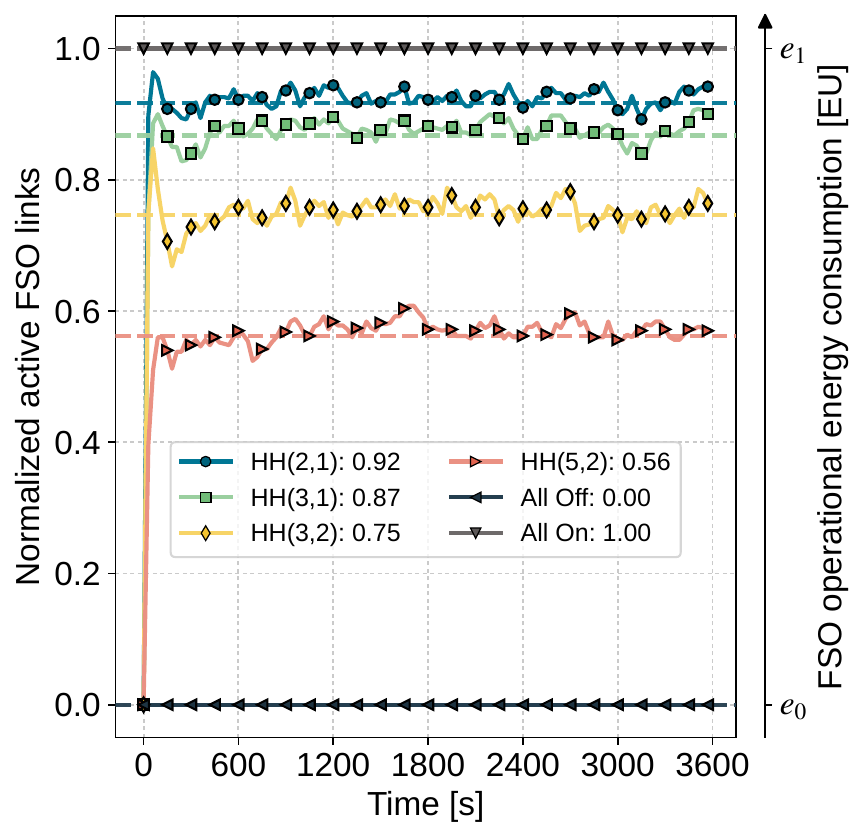}
    \caption{Normalized number of active FSO links over time, which also corresponds to the normalized FSO operational energy consumption.}
    \label{fig:results-on-time}
\end{figure}

Fig.~\ref{fig:results-coverage} presents user coverage over time (dashed lines indicate averages).
The \textit{All On} strategy achieves 81.8\% coverage, while \textit{All Off} yields 61.9\%.
HH(2,1) achieves coverage only 0.9\% below the \textit{All On} benchmark, while HH(5,2) results in a reduction of 6.7\%.
Normalizing to the achievable range (\textit{All On} minus \textit{All Off}), the HH variants attain 96.5\%, 94.0\%, 87.4\%, and 72.9\% of the \textit{All On} highest performance for threshold pairs $(2, 1)$, $(3, 1)$, $(3, 2)$, and $(5, 2)$, respectively.

\definecolor{colorRenseFlamingo}{HTML}{e1634f}
\tikzset{
    selectedAreaLabel/.style={
        text=colorRenseFlamingo,
        font=\bfseries\small,
        fill=white,
        fill opacity=0.8,
        text opacity=1,
        inner sep=2pt,
        anchor=south west
    }
}

\begin{figure}[t]
    \centering
    \begin{tikzpicture}[y=-1cm]
    \node[anchor=north west, inner sep=0] at (0,0) {\includegraphics[width=.9\columnwidth, trim={0 0.2cm 0 0}, clip]{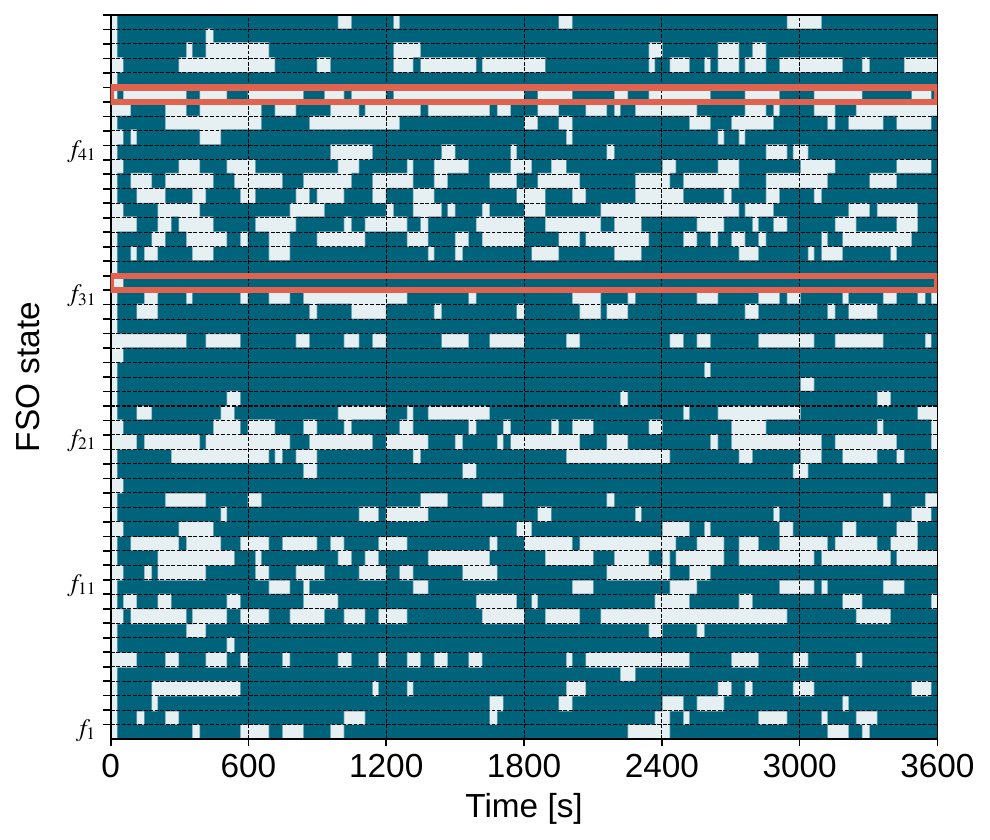} \hfill};
    \node[selectedAreaLabel] at (6.5, 0.82) {$f_{45}$};
    \node[selectedAreaLabel] at (6.5, 2.2) {$f_{32}$};
    \end{tikzpicture}%
    \caption{FSO state for HH(3,2) under traffic pattern~A.}
    \label{fig:results-heatmap}
\end{figure}
\begin{figure}[t]
    \centering
    \begin{tikzpicture}[y=-1cm]
    \node[anchor=north west, inner sep=0] at (0,0) { \includegraphics[width=.9\columnwidth, trim={0 0.2cm 0 0}, clip]{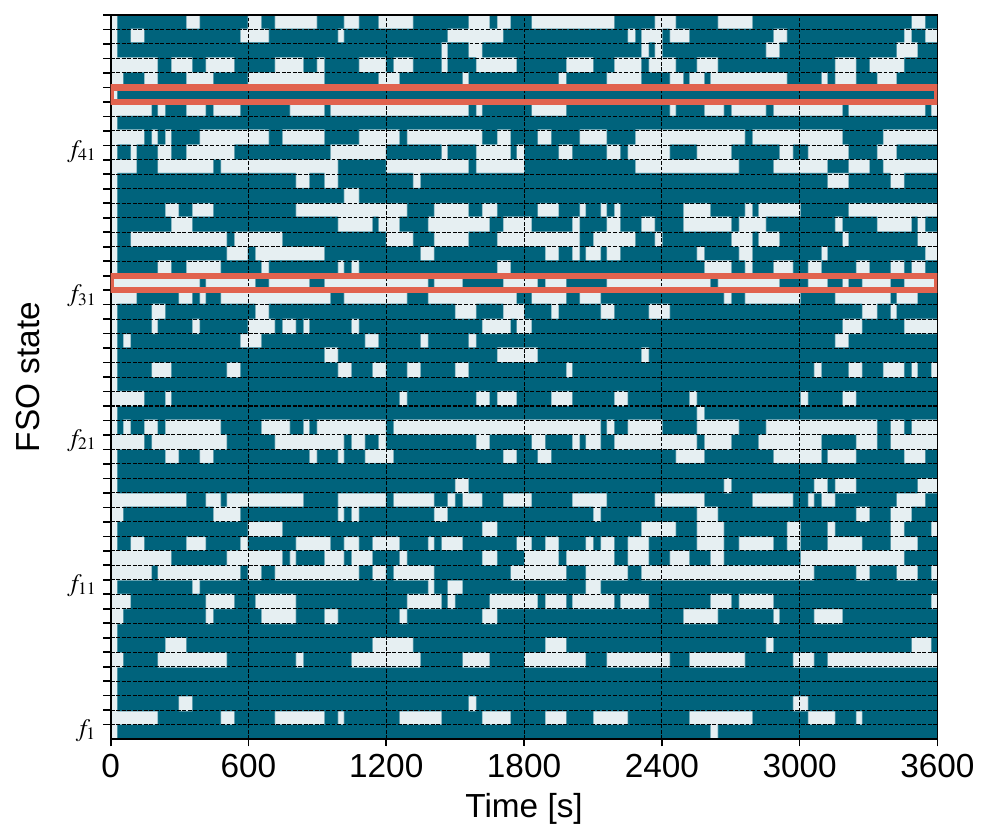} \hfill};
    \node[selectedAreaLabel] at (6.5, 0.82) {$f_{45}$};
    \node[selectedAreaLabel] at (6.5, 2.2) {$f_{32}$};
    \end{tikzpicture}%
    \caption{FSO state for HH(3,2) under traffic pattern~B.}
    \label{fig:results-heatmap-2}
\end{figure}

Fig.~\ref{fig:results-on-time} shows \ac{FSO} operational energy consumption over time.
HH(2,1) reduces energy consumption by 8\% relative to \textit{All On}, while HH(5,2) achieves a 44\% reduction.
Crucially, energy drops faster than coverage across HH configurations, confirming that substantial energy savings are attainable at the cost of smaller coverage reductions.

Figs.~\ref{fig:results-heatmap} and~\ref{fig:results-heatmap-2} show the per-\ac{SBS} \ac{FSO} state over time under two distinct traffic patterns.
Dark shading denotes active \ac{FSO}; light shading denotes inactive.
The set of \acp{SBS} benefiting from additional \ac{FSO} capacity varies both temporally and across traffic patterns.
Two representative nodes, $f_{32}$ and $f_{45}$, illustrate this clearly.
Node $f_{32}$ remains active throughout the simulation under pattern~A but is mostly inactive under pattern~B.
Conversely, $f_{45}$ is predominantly inactive under pattern~A yet active under pattern~B.
This confirms that the need for high-capacity backhaul is inherently stochastic: no static a priori assignment can match the adaptivity of dynamic closed-loop control.

\section{Conclusion}

Dynamic operation of hybrid fiber-\ac{IAB}-\ac{FSO} topologies yields meaningful energy savings under \ac{UE} mobility.
By selectively activating auxiliary \ac{FSO} links based on per-\ac{SBS} user load, our closed-loop control mechanism cuts energy by 8\% with a 0.9\% coverage loss.
Accepting a 6.7\% coverage drop raises energy savings up to 44\%.
A per-FSO-link state analysis confirms that nodes requiring high-capacity backhaul vary stochastically, highlighting the necessity of dynamic strategies over static planning.
Future work will incorporate \ac{FSO} channel-state monitoring into the control loop to maintain reliability under adverse weather.

\section{Acknowledgements}
This work has been supported by the Horizon Europe ECO-eNET project, funded by the SNS JU under grant agreement No. 101139133.


\printbibliography

\vspace{-4mm}

\end{document}